\documentclass[12pt]{iopart}
\usepackage{graphicx}
\usepackage{bm}
\usepackage{amsbsy}
\usepackage{amssymb}

\begin{document}

\title{Distances between quantum states in the tomographic-probability representation}

\author{S N Filippov$^1$ and V I Man'ko$^2$}

\address{$^1$ Moscow Institute of Physics and Technology, Moscow, Russia}

\address{$^2$ P N Lebedev Physical Institute, Moscow, Russia}

\eads{\mailto{filippovsn@gmail.com},
\mailto{manko@sci.lebedev.ru}}

\begin{abstract}
Distances between quantum states are reviewed within the framework
of the tomographic-probability representation. Tomographic
approach is based on observed probabilities and is straightforward
for data processing. Different states are distinguished by
comparing corresponding probability-distribution functions.
Fidelity as well as other distance measures are expressed in terms
of tomograms.
\end{abstract}


\maketitle

\section{\label{introduction}Introduction}

Comparing quantum states is of crucial importance for the theory
of quantum information processing. At any step of computation one
should be aware of a system state and know how much it deviates
from desired evolution. Moreover, it is necessary to compare
different states and quantify the distance between them by virtue
of experimental data. Measurements enable us to inquire some
information about system but such an information is presented in
the form of observed probabilities. For this reason one needs a
way of comparing quantum states with the help of measured
probability distributions and try not to employ the density matrix
formalism.

Probability-distribution functions enable one to reconstruct
quantum states
\cite{tombesi-manko,ber-ber,vog-ris,dodonovPLA,oman'ko-jetp}. On
the other hand, quantum states can be identified with observed
probability distributions rather than density operators or Wigner
functions \cite{oman'ko-97,mendes-physica}. According to this
representation, quantum states are associated with fair
probability-distribution functions called tomograms. The
tomographic representation of quantum states with continuous
variables (position, momentum) is introduced in
\cite{tombesi-manko} and that of states with discrete variables
(spins) is introduced in \cite{dodonovPLA,oman'ko-jetp}. Such an
approach describes not only states but can also be developed to
build the whole tomographic picture of quantum mechanics (see the
reviews \cite{sudarshan-2008,ibort}).

In this paper, we use peculiarities of the tomographic-probability
representation to express distances between spin and light states
in terms of quantum tomograms. The main idea of this consideration
is that two states are close to each other if the corresponding
probability distributions differ slightly, and are far apart from
each other if their tomograms do not match significantly. These
basic ideas were outlined and successfully applied to Fock's,
coherent, squeezed, and Schr\"{o}dinger cat states in
\cite{dodonov-m-m-wunsche,wunsche-dod-m-m}.

The paper is organized as follows.

In \Sref{spin-tomogram}, we are aimed at recalling the tomographic
representation of spin and light states and observables. Star
product of tomographic symbols is also introduced in concise
manner. In \Sref{sec::distances}, we consider conventional
distance measures which are often used in quantum information
theory. Tomographic analogues of these distances are introduced
for spins (qubits, qudits) in \Sref{subsec::spin-distance} and for
photon states in \Sref{subsec::photon-distance}. In
\Sref{conclusions}, conclusions and prospects are presented.

\section{\label{spin-tomogram}State tomograms}

To begin with, whatever system is under investigation (spin,
qubit, qudit, light, particle) it can be associated with
tomographic symbol of the form

\begin{equation}
w({\bf x}) = \Tr \left( \hat{\rho} ~ \hat{U}({\bf x}) \right),
\end{equation}

\noindent where $\hat{\rho}$ is the density operator and
$\hat{U}({\bf x})$ is a dequantizer operator depending on a
particular set of parameters ${\bf x}$.

The inverse mapping of tomographic symbols onto density operators
is

\begin{equation}
\hat{\rho} = \int \rmd {\bf x} ~w({\bf x}) ~\hat{D}({\bf x}),
\end{equation}

\noindent where $\hat{D}({\bf x})$ is a quantizer operator.

If operator $\hat{A}$ is given, it is also possible to construct
tomographic symbol of this operator. The only thing one should do
is to replace density operator $\hat{\rho}$ by $\hat{A}$. If one
knows symbols $w_A({\bf x})$ and $w_B({\bf x})$ of operators
$\hat{A}$ and $\hat{B}$, respectively, then the symbol of operator
$\hat{A}\hat{B}$ is equal to the star-product of separate
tomographic symbols. Namely,

\begin{eqnarray}
\fl w_{AB}({\bf x}) &\equiv& (w_A \star w_B) ({\bf x}) \nonumber\\
\fl &=& \int w_A({\bf x}_1) w_B({\bf x}_2) K({\bf x},{\bf
x}_1,{\bf x}_2) \rmd {\bf x},
\end{eqnarray}

\noindent where $K({\bf x},{\bf x}_1,{\bf x}_2)$ is called
tomographic star-product kernel. General and specific tomographic
star-product schemes are discussed in
\cite{man'ko-new-star-product,marmoJPA,oman'ko-star-product,patr}.
The problem of spin tomographic kernels for spins was attacked
from different perspectives, e.g., in
\cite{castanos,filip-star,filip-chebyshev}.

Particular form of quantizer and dequantizer operators depends on
the system in question. Further we epitomize these operators for
spin and light systems.

\subsection{Spin tomography}

By convention, basis states are eigenvectors $|jm\rangle$ of
angular momentum operators $\hat{J}_z$ and $\hat{\bf J}^2$. Let
$u$ be an element of group $SU(2)$ or $SU(N)$ with $N=2j+1$. We
use the following notation:

\begin{eqnarray}
{\bf x} = (m,u), \\
 \int \rmd {\bf x} = \sum \limits_{m=-j}^{j}
\frac{1}{8\pi^{2}} \int \limits_{0}^{2\pi} {\rmd}\alpha \int
\limits_{0}^{\pi} \sin\beta {\rmd}\beta \int
\limits_{0}^{2\pi}{\rmd}\gamma,
\end{eqnarray}

\noindent where the latter equation implies that $u \in SU(2)$ is
parametrized by Euler angles $\alpha$, $\beta$, and $\gamma$. Then
scanning and reconstruction procedures of spin tomography are
given by dequantizer and quantizer operators of the form
\cite{filip-star}:

\begin{eqnarray}
\fl \hat{U}({\bf x}) = u^{\dag} | m \rangle \langle m| u, \\
\fl \hat{D}({\bf x}) = (2j+1) \Big[ \hat{U}({\bf x}) &-&
\frac{1}{2}
\hat{R}_{+}(u) \hat{U}({\bf x}) \hat{R}_{-}(u) \nonumber\\
\fl &-& \frac{1}{2} \hat{R}_{-}(u) \hat{U}({\bf x}) \hat{R}_{+}(u)
\Big],
\end{eqnarray}

\noindent where

\begin{eqnarray}
\hat{R}_{+}(u)= \sum\limits_{m=-j}^{j-1} u^{\dag}|j,m+1\rangle
\langle j,m|u,\\
\hat{R}_{-}(u)= \sum\limits_{m=-j+1}^{j} u^{\dag}
|j,m-1\rangle \langle j,m| u.
\end{eqnarray}

\subsection{Photon-number tomography}

Quantum state of light is uniquely determined by the photon-number
tomogram. In this case Fock states $|n\rangle$ form a basis,
displacement operator $\hat{\mathcal{D}}(\alpha) = \exp [\alpha
\hat{a}^{\dag} - \alpha^{\ast} \hat{a}]$ plays role of unitary
matrix in the spin tomography. In other words, the notation is

\begin{equation}
{\bf x} = (n,\alpha), \qquad \int \rmd {\bf x} = \sum
\limits_{n=0}^{\infty} \int\!\!\!\!\int \frac{\rmd^2 \alpha}{\pi},
\end{equation}

\noindent and tomographic procedures are given by operators
\cite{WV,banaszek,m-m-tombesi,manko-jrlr-03}

\begin{eqnarray}
\hat{U}({\bf x}) = \hat{\mathcal{D}}^{\dag}(\alpha) | n \rangle \langle n | \hat{\mathcal{D}}(\alpha), \\
\hat{D}({\bf x}) = \frac{4}{1-s^2} \left( \frac{s-1}{s+1}
\right)^{(\hat{a}^{\dag} + \alpha^{\ast})(\hat{a} + \alpha)-n},
\end{eqnarray}

\noindent where $s$ is an arbitrary ordering parameter
\cite{cahill}.

\section{\label{sec::distances}Distances between states in view of their tomograms}

The extent to which quantum states are similar to each other is
usually expressed in terms of their density operators
$\hat{\rho}_1$ and $\hat{\rho}_2$. On the other hand, unlike
tomogram density matrix is not observed directly in experiment.
This fact makes reasonable to express basic distance measures
between states in terms of their tomograms. In this paper, we are
going to explore the following standard quantities (see, e.g.,
\cite{fuchs,nielsen,uhlmann,jozsa,teiko-mario}):

(i) Hilbert-Schmidt distance $\| \rho_1 - \rho_2 \|_{\rm HS}
\equiv \left[ \frac{1}{2} \Tr (\rho_1 - \rho_2)^2 \right]^{1/2}$;

(ii) trace distance $\frac{1}{2} \Tr | \rho_1 - \rho_2 |$, where
$|A| \equiv \sqrt{A^{\dag} A}$;

(iii) fidelity $F(\rho_1,\rho_2)= \Tr \left[ \sqrt{\rho_1} \rho_2
\sqrt{\rho_1} \right]^{1/2}$;

(iv) operator norm $\| \rho_1 - \rho_2 \| =
\sup\limits_{\|\psi\|=1} \| (\rho_1 - \rho_2)\psi \|$.

If the reconstruction procedure is given, one is tempted to use
quantizer operator $\hat{D}({\bf x})$ instead of density operator
whenever it is possible. As far as distance measures are
concerned, such a method does not give us any advantage as
compared to density operators. Here we develop an alternative
approach which is based on probability-distribution functions
only, is straightforward for computation and sheds some light on
relation between measures for classical and quantum information.

Both spin and photon number tomogram can be considered as
probability vectors due to normalization conditions
$\sum_{m=-j}^{j}w(m,u) = 1$ and $\sum_{n=0}^{\infty} w(n,\alpha) =
1$. Indeed, splitting different numbers $m$ and $n$, we obtain

\begin{eqnarray}
\fl \overrightarrow{w}(u) = \left(%
\begin{array}{cccc}
  w(j,u) & w(j-1,u) & \cdots & w(-j,u) \\
\end{array}%
\right)^{\tr}, \\
\fl \overrightarrow{w}(\alpha) = \left(%
\begin{array}{cccc}
  w(0,\alpha) & w(1,\alpha) & w(2,\alpha) & \cdots \\
\end{array}%
\right)^{\tr}.
\end{eqnarray}

We will discuss spin systems (qubits, qudits) first and then we
will also pay attention to the photon-number tomogram.

\subsection{\label{subsec::spin-distance}Distances in terms of spin tomograms}

In the paper \cite{filip-geometrical}, it is shown that the
Hilbert-Schmidt distance between qubit ($j=1/2$) states is equal
to the maximal possible Euclidean distance between corresponding
tomographic-probability vectors with respect to rotations in
Hilbert space. Here, we extend this claim to higher dimensions.

{\bf Proposition.} For an arbitrary qudit states $\rho_1$ and
$\rho_2$ the following relation takes place ($N=2j+1$):

\begin{eqnarray}
\fl \| \rho_1 - \rho_2 \|_{\rm HS} \nonumber\\
\fl = \max\limits_{u \in SU(N)} \left[ \frac{1}{2} \sum_{m=-j}^{j}
\Big( w_1(m,u) - w_2(m,u) \Big)^2 \right]^{1/2}.
\end{eqnarray}

{\bf Proof.}

Tomogram of any Hermitian operator $\hat{A}$ can be rewritten in
terms of its eigenvalues ${A_l}$, $l=1,\dots,N$ and unitary matrix
$u_A$ composed of its eigenvectors as follows \cite{ibort}:

\begin{eqnarray}
\label{bistochastic-tomogram} \fl
\overrightarrow{w}_A (u) 
= \left(%
\begin{array}{ccc}
  |(uu_A)_{11}|^2 & ... & |(uu_A)_{1N}|^2 \\
  ... & ... & ... \\
  |(uu_A)_{N1}|^2 & ... & |(uu_A)_{NN}|^2 \\
\end{array}%
\right) \left(%
\begin{array}{c}
  A_1 \\
  ... \\
  A_N \\
\end{array}%
\right).
\end{eqnarray}

%
%

\noindent By $M$ denote $N\times N$ matrix in
(\ref{bistochastic-tomogram}). Note that $M$ is bistochastic since
all rows and all columns sum to 1. It means that each component of
vector $\overrightarrow{w}_A(u)$ is a convex sum of eigenvalues
$A_k$. Moreover, sum of all components of vector
$\overrightarrow{w}_A(u)$ equals sum of $A_k$. Due to these facts
the maximal value of quantity

\begin{equation}
\label{M2-function} \fl \Big( \overrightarrow{w}_A (u),
\overrightarrow{w}_A (u) \Big) \equiv \sum_{k=1}^{N} w_{A~k}^2 (u)
= \sum_{k=1}^{N} \left( \sum_{l=1}^{N} M_{kl}A_l \right)^2
\end{equation}

\noindent is achieved when $M$ is identity matrix or,
equivalently, $u=u_A^{\dag}$. In other words,

\begin{eqnarray}
\fl \max\limits_{u \in SU(N)} \Big( \overrightarrow{w}_A (u),
\overrightarrow{w}_A (u) \Big) &=& \Big( \overrightarrow{w}_A
(u_A),
\overrightarrow{w}_A (u_A) \Big) \nonumber\\
\fl &=& \sum_{k=1}^{N} A_k^2 = \Tr A^2.
\end{eqnarray}

\noindent If we replace $A$ by $(\rho_1 - \rho_2)$ and recall $\|
\rho_1 - \rho_2 \|_{\rm HS} = [\frac{1}{2} \Tr (\rho_1 -
\rho_2)^2]^{1/2}$, we obtain the statement of the proposition.

$\Box$

In addition to this relation it is worth emphasizing that
rotations by elements $u \in SU(2) $ result in smaller distance

\begin{eqnarray}
\fl 0 &\le& \max\limits_{u \in SU(2)} \left[ \frac{1}{2}
\sum_{m=-j}^{j} \Big(
w_1(m,u) - w_2(m,u) \Big)^2 \right]^{1/2} \nonumber\\
\fl &\le& \| \rho_1 - \rho_2 \|_{\rm HS},
\end{eqnarray}

\noindent however, it can also serve as measure of distance
because left inequality becomes equality iff $\rho_1 \equiv
\rho_2$.

{\bf Proposition.} Trace distance between spin states $\rho_1$ and
$\rho_2$ is expressed in terms of tomograms as follows:

\begin{eqnarray}
\label{trace-distance-tom} \fl \frac{1}{2}\Tr | \rho_1 - \rho_2 |
\nonumber\\
\fl = \max\limits_{u\in SU(N)} \left[ \frac{1}{2} \sum_{m=-j}^{j}
|w_1(m,u) - w_2(m,u)| \right]
\end{eqnarray}

{\bf Proof.} Arguing as above, we obtain the similar equation for
operator $A=\rho_1-\rho_2$. Namely,

\begin{equation}
\sum_{k=1}^{N} \left| w_{A~k} (u) \right| = \sum_{k=1}^{N} \left|
\sum_{l=1}^{N} M_{kl}A_l \right|.
\end{equation}

\noindent This function of variables $M_{kl}$ has the same
properties as function (\ref{M2-function}) and is harmonic. For
these reasons it also achieves its maximal value at the boundary
determined by bistochastic matrix $M$, with the maximum being
equal to

\begin{equation}
\sum_{k=1}^{N} |A_k| = \Tr |A|.
\end{equation}

$\Box$

In other words, trace distance is equal to the maximal possible
Kolmogorov distance between tomographic-probability distributions.
Note that maximum is attained by the same element $\tilde{u}$ of
$SU(N)$ as in case of Hilbert-Schmidt distance.

Let us now consider fidelity.

{\bf Proposition.} Tomographic-probability version of fidelity
reads

\begin{eqnarray}
\fl \Tr \left[ \sqrt{\rho_1} \rho_2 \sqrt{\rho_1} \right]^{1/2}
\nonumber\\
\fl = \min\limits_{u \in SU(N)} \left[ \sum_{m=-j}^{j}
\sqrt{w_1(m,u) ~ w_2(m,u)} \right]
\end{eqnarray}

{\bf Proof.} To prove this Proposition one can follow, step by
step, proof of the known formula (see, e.g., \cite{nielsen})

\begin{equation}
\label{fidelity-min} F(\rho_1,\rho_2) = \min\limits_{\{ E_k \}}
\sum_{k} \sqrt{\Tr(\rho_1 E_k) \Tr(\rho_2 E_k)},
\end{equation}

\noindent where minimum is over all positive operator valued
measures $\{ E_k \}$. To prove that the minimum is achieved the
authors \cite{nielsen} use effects $E_k$ in the form of projectors
$|\varphi_k\rangle\langle\varphi_k|$. In our case we can employ
$E_m = \sqrt{E_m} = u|m\rangle\langle m| u^{\dag}$ with matrix $u$
such that $|\varphi_k\rangle = u |k\rangle$.

$\Box$

It is worth noting that fidelity is the minimal value of
Bhattacharyya coefficient \cite{bhattacharyya} (see also the
review \cite{teiko-mario}) of two tomographic-probability
distributions.

Another way to consider fidelity is to use symmetric form
$F=\Tr\left|\sqrt{\rho_1}\sqrt{\rho_2}\right|$ and tomographic
symbols of operators $\sqrt{\rho_1}$ and $\sqrt{\rho_2}$. The
question arises itself how to express tomogram $w_{\sqrt{\rho}}
(m,u)$ of positive operator $\sqrt{\rho}$ if we know tomogram of
$w_{\rho} (m,u)$ of state $\rho$. Using spectral decomposition of
density operator

\begin{equation}
\label{rho-sp-decomposition} \rho = \sum_{m'=-j}^{j} \rho_{m'}
~u_{\rho} |m'\rangle \langle m'| u_{\rho}^{\dag},
\end{equation}

\noindent it is easy to express tomogram

\begin{equation}
\fl w_{\rho} (m,u) = \langle m| u \rho u^{\dag}| m\rangle =
\sum_{m'=-j}^{j} \rho_{m'} \left| \langle m | u u_{\rho}
|m'\rangle \right|^2.
\end{equation}

\noindent From this it follows that
$\rho_{m'}=w_{\rho}(m',u_{\rho}^{\dag})$. Spectral decomposition
of operator $\sqrt{\rho}$ is obtained from
(\ref{rho-sp-decomposition}) by replacing $\rho_{m'}$ by
$\sqrt{\rho_{m'}}$. Then we have

\begin{equation}
\fl w_{\sqrt{\rho}} (m,u) = \sum_{m'=-j}^{j} \sqrt{w_{\rho}
(m',u_{\rho}^{\dag})} ~ \left| \langle m | u u_{\rho} | m' \rangle
\right|^2.
\end{equation}

Tomographic symbol of operator $\sqrt{\rho_1}\sqrt{\rho_2}$ is the
star product of corresponding symbols. Kernel of this star-product
for qudits is calculated, e.g., in
\cite{castanos,filip-chebyshev}. Taking advantage of
(\ref{trace-distance-tom}) we obtain

\begin{equation}
\label{fidelity-star} \fl \Tr |\sqrt{\rho_1}\sqrt{\rho_2}| =
\max\limits_{u\in SU(N)} \left[ \sum_{m=-j}^{j}
\left|(w_{\sqrt{\rho_1}} \star w_{\sqrt{\rho_2}})(m,u) \right|
\right]
\end{equation}

If we compare (\ref{fidelity-min}) and (\ref{fidelity-star}), we
reveal new properties of tomograms. In fact,

\begin{eqnarray}
\min\limits_{u \in SU(N)} \left[ \sum_{m=-j}^{j}
\sqrt{w_{\rho_1}(m,u) ~
w_{\rho_2}(m,u)} \right] \nonumber \\
= \max\limits_{u\in SU(N)} \left[ \sum_{m=-j}^{j}
\left|(w_{\sqrt{\rho_1}} \star w_{\sqrt{\rho_2}})(m,u) \right|
\right]
\end{eqnarray}

As far as operator norm $\| \rho_1 - \rho_2 \|$ is concerned, it
is equal to the maximal eigenvalue of operator $| \rho_1 - \rho_2
|$. Consequently, this norm is

\begin{equation}
\fl \| \rho_1 - \rho_2 \| = \max\limits_{m=-j,\dots,j; ~u\in
SU(N)} |w_1(m,u) - w_2(m,u)|.
\end{equation}

\subsection{\label{subsec::photon-distance}Distances in terms of photon number tomograms}

Using analogy of spin and photon number tomogram, one can readily
extend these results to the Hilbert-Schmidt distance between light
states. For instance, Hilbert-Schmidt distance is

\begin{eqnarray}
\fl \| \rho_1 - \rho_2 \|_{\rm HS} = \max\limits_{\mathcal{D} \in
SU(\infty)} \left[ \frac{1}{2} \sum_{n=0}^{\infty} \langle n |
\mathcal{D} (\rho_1 -\rho_2) \mathcal{D}^{\dag} | n \rangle^2
\right]^{1/2}.
\end{eqnarray}

So as not to resort to group $SU(\infty)$ we will only formulate
inequalities in term of conventional photon number tomograms:

\begin{eqnarray}
\fl 0 \le \max\limits_{\alpha \in \mathbb{C}} \left[ \frac{1}{2}
\sum_{n=0}^{\infty} \Big( w_1(n,\alpha) - w_2(n,\alpha) \Big)^2
\right]^{1/2} \nonumber\\
\fl \quad \le \| \rho_1 - \rho_2 \|_{\rm HS},\nonumber\\
\fl 0 \le \max\limits_{\alpha \in \mathbb{C}} \left[ \frac{1}{2}
\sum_{n=0}^{\infty} \left| w_1(n,\alpha) - w_2(n,\alpha) \right|
\right] \le \frac{1}{2} \Tr| \rho_1 - \rho_2 |,\nonumber\\
\fl 0 \le F(\rho_1,\rho_2) \le \min\limits_{\alpha \in \mathbb{C}}
\left[ \sum_{n=0}^{\infty} \sqrt{w_1(n,\alpha) ~ w_2(n,\alpha)}
\right],\nonumber\\
\fl 0 \le \max\limits_{n \in \{0\} \cup \mathbb{N}, ~\alpha \in
\mathbb{C}} \left| w_1(n,\alpha) - w_2(n,\alpha) \right| \le
\|\rho_1-\rho_2\|.\nonumber
\end{eqnarray}

\section{\label{conclusions}Conclusions}

To conclude we summarize the main results of the paper.

Conventional distance measures between quantum states and
fidelity, which are usually formulated for density matrices, are
expressed in terms of quantum tomograms. It is demonstrated that
Hilbert-Schmidt distance is related to maximal Euclidean distance
of tomographic-probability vectors, trace distance is related to
maximal Kolmogorov distance of tomograms, fidelity is related to
minimal Bhattacharyya coefficient, and operator norm is related to
maximum of residual tomographic symbol. Analyzing photon number
tomography, these results are also extended to the case of
infinite Hilbert space of Fock states and formulated in the form
of inequalities. We believe the introduced quantities to be used
as an alternative to distance measures based on density matrices.
Interesting problem for further consideration is to develop
analogues approach for continuous variables quantum systems.

\ack This study was partially supported by the Russian Foundation
for Basic Research under Project Nos. 07-02-00598, 08-02-90300,
and 09-02-00142. SNF thanks the Ministry of Education and Science
of the Russian Federation and the Federal Education Agency for
support under Project No. 2.1.1/5909. The authors are grateful to
the Organizers of the Sixteenth Central European Workshop on
Quantum Optics (Turku, Finland, May 23-27, 2009) for invitation
and kind hospitality. SNF would like to express his gratitude to
the Organizing Committee of the Conference and especially to
Professor Kalle-Antti Suominen for financial support. SNF thanks
the Russian Foundation for Basic Research for travel grant No.
09-02-09240.

\end{document}